\documentstyle[psfig,aps,twocolumn]{revtex}

\def\beq{\begin{equation}}
\def\eeq{\end{equation}}
\def\bea{\begin{eqnarray}}
\def\eea{\end{eqnarray}}
\newcommand{\sla}{\!\!\!\!/ \,}

\begin{document}


\title{Non-Equilibrium HTL Resummation} 
\author{M.E. Carrington$^1$, Hou Defu$^2$, and M.H. 
Thoma$^{3,4}$}
\address{$^1$Department of Physics, Brandon University, Brandon, Manitoba, Canada R7A 6A9 \\
$^2$Institute of Particle Physics, Huazhong Normal University, 
430070 Wuhan, China\\
$^3$Institut f\"ur Theoretische Physik, Universit\"at Giessen,
D-35392 Giessen, Germany\\
$^4$Institut f\"ur Theoretische Physik, Universit\"at Regensburg, 
D-93040 Regensburg, Germany}
\maketitle

\begin{abstract}
We present an extension of the HTL resummation technique to non-equilibrium 
situations, starting from the real time formalism in the Keldysh 
representation. As an example we calculate the HTL photon self energy,
from which we derive the resummed photon propagator. This propagator
is applied to the computation of the interaction rate of a hard electron
in a non-equilibrium QED plasma. In particular we show the absence of
of pinch singularities in these quantities. Finally we show that the Ward
identities relating the HTL electron self energy to the HTL electron-photon
vertex hold as in equilibrium.  
\end{abstract}

\section{Introduction}

Perturbative QCD at finite temperature and density can be used to describe 
properties of an equilibrated quark-gluon plasma (QGP). However, using
only bare propagators and vertices infrared divergent and gauge dependent 
result are obtained. These problems can be partly avoided by using
the hard thermal loop (HTL) resummation technique \cite{ref1}. In this
way consistent results, which are complete to leading order in the
coupling constant and gauge independent, are found. At the same time 
the infrared behavior is improved by the presence of effective masses in
the HTL propagators. This method has been applied to various quantities
of the QGP, such as damping rates of quarks and gluons, energy loss
of energetic partons, thermalization times, viscosity, and photon and
dilepton production \cite{ref2}.

The HTL method has been derived within the imaginary time formalism,
which is restricted to equilibrium situations. However, in relativistic 
heavy ion collisions the formation of a pre-equilibrium parton gas has to 
be expected. Indeed it is not clear whether a complete thermal and chemical
equilibrium will be reached in the evolution of the fireball at all
\cite{ref3}. Therefore it is desirable to extend the HTL resummation 
technique to non-equilibrium. For this purpose we have to start from
the real time formalism (RTF) \cite{ref4}. Similar investigations
have been performed by Baier et al. \cite{ref5}.

In the RTF formalism propagators are given by $2\times 2$-matrices 
\cite{ref6}. For example the scalar propagator reads
\bea
D(K) & = & \left (\begin{array}{cc} \frac{1}{K^2-m^2+i\epsilon} & 0\\
                             0 & \frac{-1}{K^2-m^2-i\epsilon}\\
            \end{array} \right ) -  2\pi i\, \delta (K^2-m^2)\nonumber \\
&& \times \left (\begin{array}{cc}
f_B & \theta (-k_0)+f_B\\
\theta (k_0)+f_B & f_B \\ \end{array} \right ). 
\label{e1}
\eea
Here we use the following notation: $K=(k_0,{\bf k})$, $k=|{\bf k}|$.
In equilibrium the distribution function $f_B$ is given by the
Bose distribution $n_B(k_0)=1/[\exp(|k_0|/T)-1]$. In non-equilibrium
the equilibrium distributions should be replaced by Wigner functions 
$f_B=f_B(|k_0|, {\vec k}, x)$ \cite{ref7}.

A particular useful representation of the propagators in the RTF
is the Keldysh representation, where retarded, advanced, and symmetric
propagators are defined as linear combinations of the four components
of (\ref{e1}), of which only three are independent \cite{ref8}:
\bea
D_R & = & D_{11}-D_{12},\nonumber \\
D_A & = & D_{11}-D_{21},\\
D_F & = & D_{11}+D_{22}.\nonumber
\label{e2}
\eea
For example, the bare scalar propagator in the Keldysh representation
is given by
\bea
D_R(K) & = & \frac{1}{K^2-m^2+i\, \mbox{sgn}(k_0) \epsilon},\nonumber \\
D_A(K) & = & \frac{1}{K^2-m^2-i\, \mbox{sgn}(k_0) \epsilon}, \\
D_F(K) & = & -2\pi i\, (1+2f_B)\, \delta (K^2-m^2).\nonumber 
\label{e3}
\eea
The distribution functions only appear in the symmetric component $D_F$,
which is of particular advantage for the HTL diagrams where the terms
containing distribution functions dominate.  

\section{HTL photon self energy}

Before we are going to calculate HTL's in non-equilibrium we would like to 
note here that we do not aim to describe the thermalization of a partonic
fireball starting far from equilibrium. As discussed in the literature
(see e.g. \cite{ref9}) this problem requires the use of non-perturbative
methods beyond the HTL perturbation theory. We have situations in mind
where we can use quasistatic distributions, i.e., where the equilibration
is slow compared to the process under consideration. Such a situation
might be the case for the chemical equilibration of the parton gas
in relativistic heavy ion collisions after a thermal equilibrium has 
been achieved rapidly \cite{ref10}. Also anisotropic momentum distributions 
taking account of the longitudinal and transverse expansion might be 
considered \cite{ref11}.
At least as long as we are close to equilibrium the use of quasistatic
distributions within the HTL method, based on the weak coupling
limit $g\ll 1$, is justified by the fact that the equilibration time $\tau 
\sim (g^2T)^{-1}$ is large compared to the HTL time scale $(gT)^{-1}$.

We have chosen a non-equilibrium QED plasma as an example for the use
of the HTL method as the computation of the HTL photon self energy
is much easier than the one of the HTL gluon self energy. After all,
the results for these self energies differ only by a constant color 
and flavor factor.  

The HTL photon self energy is given by the one-loop diagram shown in Fig.1,
where the internal momenta are hard compared to the external. In 
equilibrium this can be achieved in the weak coupling limit by considering 
internal momenta of the order of the temperature $T$ or larger and 
external of the order $gT$ or smaller. In non-equilibrium we might
replace the temperature by the average momentum following from the
distribution function under consideration. 

\begin{figure}
\centerline{\psfig{figure=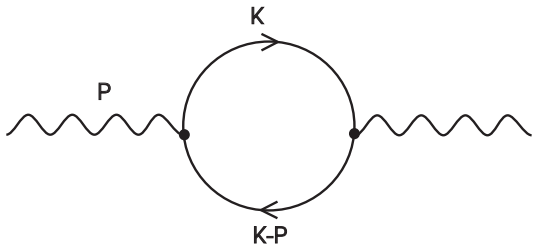,width=7cm}}
\caption{HTL photon self energy}
\end{figure}

Here we restrict ourselves to the longitudinal component of the self energy,
i.e. $\Pi_L=\Pi_{00}$. Using the Keldysh representation and $S(K)=
K\sla \> \Delta (K)$ for the fermion propagators, we find
\bea
&& \Pi _R^L(P)=-2ie^2\int \frac{d^4K}{(2\pi )^4} (q_0k_0+{\bf q}\cdot {\bf k})
\nonumber \\ && \times \biggl [\Delta _F(K-P)\Delta _R(K)+\Delta _A(K-P)
\Delta _F(K) \biggr ].
\label{e4}
\eea
In the HTL approximation ($K\gg P$) the retarded ($\Pi_R=\Pi_{11}+\Pi_{12}$)
and advanced ($\Pi_A=\Pi_{11}+\Pi_{21}$) self energies are given then by
\beq
\Pi_{R,A}^L(P)=-3 m_\gamma^2 \left (1-\frac{p_0}{2 p}\ln \frac{p_0+p\pm 
i\epsilon} {p_0-p\pm i\epsilon} \right),
\label{e5}
\eeq
where the effective photon mass is given by an integral over the 
non-equilibrium Fermi distribution $f_F$
\beq
m_\gamma ^2=\frac{4e^2}{3\pi ^2}\int _0^\infty dk\, k\, f_F(k,x).
\label{e6}
\eeq
In equilibrium this expression reduces to $m_\gamma ^2=e^2T^2/9$.

The symmetric self energy ($\Pi_F=\Pi_{11}+\Pi_{22}$) in the HTL limit
can written as
\beq
\Pi _F^L(P)=2iA\, \frac{Im\, \Pi _R^L(P)}{p_0},
\label{e7}
\eeq
where the constant $A$ is given by
\beq
A=\frac{\int _0^\infty dk\, k^2\, f_F(k,x)\, [1-f_F(k,x)]}{\int _0^\infty
dk\, k\, f_F(k,x)}.
\label{e8}
\eeq
In equilibrium this constant reduces to $A=2T$.

\section{Resummed photon propagator}

The resummed photon propagator is constructed from the Schwinger-Dyson
equation shown in Fig.2. 

\begin{figure}
\centerline{\psfig{figure=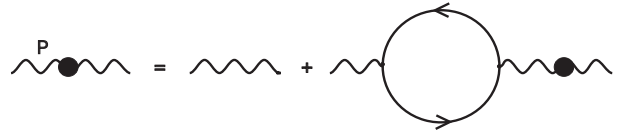,width=8cm}}
\caption{Resummed photon propagator}
\end{figure}

The Schwinger-Dyson equation for the resummed retarded and advanced  
propagators for longitudinal photons reads
\beq
{D^*}_{R,A}^L=D_{R,A}^L+D_{R,A}^L\Pi _{R,A}^L{D^*}_{R,A}^L
\label{e9}
\eeq
leading to
\beq
{D^*}_{R,A}^L(P)=\left [p^2+3m_\gamma ^2\left 
(1-\frac {p_0}{2p}\ln \frac{p_0+p\pm i\epsilon}{p_0-p\pm i\epsilon}\right )
\right ]^{-1},
\label{e10}
\eeq
where the HTL photon self energy has been used.
Apart from $m_\gamma $, defined by (\ref{e6}), this expression is identical 
to the equilibrium one.

The Schwinger-Dyson equation for the resummed symmetric propagator
is given by
\beq
{D^*}_{F}^L=D_{F}^L+D_{R}^L\Pi _R^L{D^*}_{F}^L+D_F^L\Pi _{A}^L
{D^*}_{A}^L+D_{R}^L\Pi _{F}^L{D^*}_{A}^L.
\label{e11}
\eeq
This equation is solved by the following ansatz:
\bea 
&& {D^*}_{F}^L(P) = (1+2f_B)\, \mbox{sgn}(p_0)\, 
[{D^*}_{R}^L(P)-{D^*}_{A}^L(P)] \nonumber \\
&&\; \; \; \; \; \;  +\{\Pi _F^L(P)-(1+2f_B)\, \mbox{sgn}(p_0)\, 
[\Pi _R^L(P)-\Pi _A^L(P)]\}\nonumber \\
&&\; \; \; \; \; \; \times {D^*}_{R}^L(P)\, {D^*}_A^L(P).
\label{e12}
\eea    
The product of a retarded and an advanced propagator at the same momentum 
$P$ may lead to a pinch singularity \cite{ref6}. In equilibrium, however, 
the term in the curly brackets vanishes as a consequence
of the Kubo-Martin-Schwinger boundary condition or the principle of
detailed balance. Then the full symmetric propagator is given by the
first line of (\ref{e12}), which is free of possible pinch problems
 and
reflects the dissipation-fluctuation theorem \cite{ref12}.

The following relations can easily be shown to hold in general:
\bea
&& {D^*}_R^L(P){D^*}_A^L(P)=\frac{{D^*}_R^L(P)-{D^*}_A^L(P)}{2i\, Im\, 
\Pi _R^L(P)},\nonumber \\
&& \Pi _R^L(P)-\Pi _A^L(P) =2i Im\, \Pi _R^L(P).
\label{e13}
\eea
Together with the expression (\ref{e7}) for the symmetric self energy
we find from these relations that the symmetric non-equilibrium HTL 
resummed propagator can be written as
\beq
{D^*}_F^L(P)=\frac{A}{p_0}\> [{D^*}_R^L(P)-{D^*}_A^L(P)].
\label{e14}
\eeq
Hence the symmetric propagator does not suffer from
a pinch singularity even in non-equilibrium. This observation also
holds for $p_0^2-p^2>0$, where $Im\, \Pi _R \sim \epsilon $ as
following from (\ref{e5}), because of the cancellation of $Im\, \Pi_R$
according to (\ref{e7}), (\ref{e12}), and (\ref{e13}). 

\section{Electron interaction rate}

As an example for the application of the non-equilibrium HTL resummation
technique we consider the interaction or damping rate of an energetic 
electron in a non-equilibrium QED plasma. This quantity is one of the mostly 
studied within the HTL method. (For references see \cite{ref2}.)
The HTL method does not lead to an infrared finite result in this
case since there is not enough screening in the transverse part of the
HTL photon propagator. After all we want to study this example, since 
we are only interested in the comparison between the equilibrium and the 
non-equilibrium case.

The interaction rate is defined by \cite{ref13}
\beq
\Gamma (p)=-\frac{1}{2p}\, [1-f_F(p,x)]\> tr\, [P\sla \, Im \, 
\Sigma _R(p_0=p,{\bf p})],
\label{e15}
\eeq
where $\Sigma _R$ is the retarded electron self energy. Using only bare 
propagators the lowest order contribution to $Im\, \Sigma _R$ comes from
the two-loop diagram of Fig.3.  

\begin{figure}
\centerline{\psfig{figure=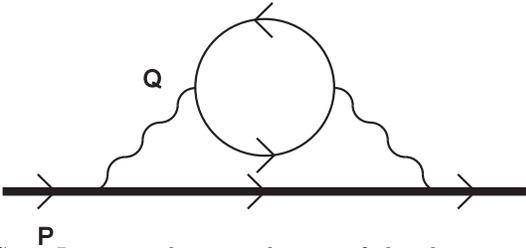,width=7cm}}
\nobreak
\caption{Lowest order contribution of the electron self energy to the 
electron interaction rate within naive perturbation theory}
\end{figure}

This diagram contains a dangerous pinch term coming from the product
of a retarded and an advanced bare propagator at the same momentum $Q$
\cite{ref14}:
\bea
D^L_R(Q) D^L_A(Q) & = & \frac{1}{Q^2 + i {\rm sgn}(q_0)\epsilon} \,\, 
\frac{1}{Q^2 - i {\rm sgn} (q_0)\epsilon}\nonumber \\
& \rightarrow & [\delta(Q^2)]^2.
\label{e16}
\eea
In equilibrium this term vanishes due to detailed balance. 

Applying the HTL method to the electron interaction rate results in 
using the one-loop diagram of Fig.4 containing a resummed photon propagator
for the electron self energy \cite{ref2,ref13}. Owing to the non-zero
imaginary part of the HTL photon self energy appearing in the
resummed photon propagator this diagram exhibits an imaginary part.
Since there is no pinch singularity in the HTL resummed photon propagator
this diagram is also free of pinch singularities. In other words,
the use of the HTL method to lowest order reduces two-loop diagrams,
containing possible pinch singularities, to one-loop diagrams
with resummed propagators, where pinch terms are absent.

\begin{figure}
\centerline{\psfig{figure=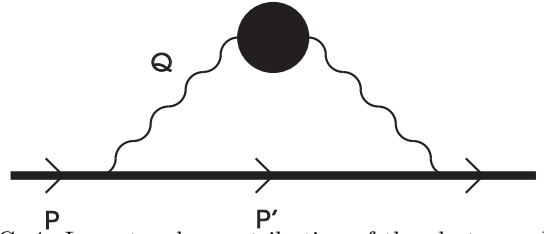,width=7cm}}
\caption{Lowest order contribution of the electron self energy to the 
electron interaction rate within the HTL method}
\end{figure}

Following the equilibrium calculation \cite{ref2,ref13}, we end up with 
the following result for the non-equilibrium interaction rate of an energetic
electron:
\beq
\Gamma _{neq}(p)\simeq \frac{e^2A}{4\pi }\> \ln \frac{const}{e}.
\label{e17}
\eeq
The logarithm in this expression reflects the presence of an infrared
singularity, which has been regularized by assuming an infrared cutoff
of the order $e^2T$. The constant under the logarithm cannot be
calculated within the HTL method. As expected this result reduces to the 
equilibrium one \cite{ref2}, if we replace $A$ by $2T$.. 

Pinch terms may arise, however, in higher order contributions of the
HTL perturbation theory, such as the one shown in Fig.5, where
the product of a retarded HTL and an advanced HTL propagator
at the same momentum $Q$ shows up. 

Resumming the photon self energy $\bar \Pi$, which contains
HTL electron propagators and HTL vertices, leads to a dressed
propagator beyond the HTL resummation. The symmetric component
of this propagator reads

\bea
&& {D^{**}}_{F}^L(Q)= \frac{A}{q_0}\, [{D^{**}}_{R}^L(Q)-{D^{**}}_{A}^L(Q)]
\\
&& +\{\bar{\Pi} _F^L(Q)-\frac{A}{q_0}\, [\bar{\Pi} _R^L(Q)-\bar{\Pi} _A^L(Q)]
\} \, {D^{**}}_{R}^L(Q)\, {D^{**}}_A^L(Q).\nonumber
\label{e18}
\eea  

If $\bar \Pi_F$ is proportional to $Im\, \bar \Pi_R$ as in (\ref{e7}),
this propagator would not contain a pinch term.

\begin{figure}
\centerline{\psfig{figure=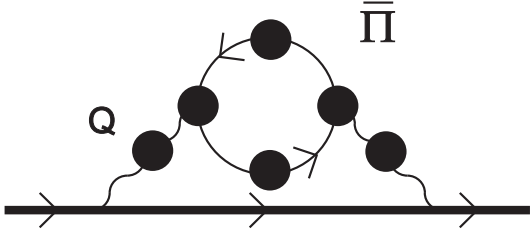,width=7cm}}
\caption{Photon self energy $\bar \Pi $}
\end{figure}
 
\section{Ward identities}

Finally we will show that the HTL electron self energy and the HTL
electron-photon vertex fulfill the same Ward identities in non-equilibrium 
as in equilibrium. For this purpose we write the one-particle-irreducible
vertex functions as \cite{ref12,ref15}
\bea
\Gamma_R &=& \frac{1}{2}(\Gamma_{111} + \Gamma_{112} 
+\Gamma_{211} + \Gamma_{212} \nonumber \\
&& \; \; \; \; - \Gamma_{121} - \Gamma_{122} 
- \Gamma_{221} - \Gamma_{222}),\nonumber \\
\Gamma_{Ri} &=& \frac{1}{2}(\Gamma_{111} + \Gamma_{112} +
 \Gamma_{121} + \Gamma_{122} \nonumber \\
&& \; \; \; \; - \Gamma_{211} -
 \Gamma_{212} -\Gamma_{221} - \Gamma_{222}),\nonumber \\
\Gamma_{Ro} &=& \frac{1}{2}(\Gamma_{111} + \Gamma_{121} +
 \Gamma_{211} + \Gamma_{221}\nonumber \\
&& \; \; \; \; - \Gamma_{112} - \Gamma_{122} -
 \Gamma_{212} - \Gamma_{222}).
\label{e19}
\eea
Here only the three retarded combinations of the seven independent components 
of the vertex in the Keldysh representation are shown. Using the HTL
approximation for the vertices and the electron self energy, we
can show the following Ward identities (without performing the loop
integrations explicitly)\cite{ref16}:
\bea
K_\phi \Gamma^\phi_{R}(-Q,P,K)& = & -ie(\Sigma_A(P) - \Sigma_R(Q)),\nonumber
\\
K_\phi \Gamma^\phi_{Ri}(-Q,P,K)& = & -ie(\Sigma_A(P) - \Sigma_A(Q)),\nonumber 
\\
K_\phi \Gamma^\phi_{Ro}(-Q,P,K)& = & -ie(\Sigma_R(P) - \Sigma_R(Q)),
\label{e20} 
\eea
which hold in equilibrium as well as non-equilibrium.

\section{Conclusions}

In the present work we have extended the HTL resummation technique
to non-equilibrium situations assuming quasistatic distribution functions.
We are not able to treat the evolution of the system towards equilibrium,
but can describe properties of a relativistic non-equilibrium plasma,
where the equilibration is slow compared to the process under 
investigation. Such a situation might be realized for example in
a chemically non-equilibrated QPG produced in the fireball of a relativistic
heavy ion collision.

Starting from the RTF we have demonstrated the usefulness of the Keldysh
representation for the HTL method in equilibrium as well as non-equilibrium.

The non-equilibrium HTL resummation technique differs from the equilibrium one
only by two aspects: 

1. There are new effective masses defined by integrals over the 
non-equilibrium distributions appearing in the HTL self energies.

2. There is a constant factor $A$, also defined by integrals over the
distributions, in the symmetric self energy and the resummed symmetric
propagator, which reduces to $2T$ in equilibrium. 
 
We have shown that there are no pinch terms in the HTL resummed propagators.
Therefore pinch singularities are absent in quantities calculated from
one-loop HTL self energies, such as the electron interaction rate. A similar 
result was found in the case of the photon production rate \cite{ref5}
and the collisional energy loss of energetic quarks \cite{ref17}.

Beyond leading order in the HTL perturbation theory we expect also
the absence of pinch terms using a higher order resummation scheme.

Finally we have given the Ward identities for the HTL electron
self energy and the HTL electron-photon vertex, which hold in equilibrium 
as well as non-equilibrium.

\end{document}